\documentclass{PoS}

\newcommand {\su}[1]{\mathrm{SU}(#1)}
\newcommand {\so}[1]{\mathrm{SO}(#1)}
\newcommand {\gu}[1]{\mathrm{U}(#1)}

\title{Phenomenology of a Composite Higgs Model}

\ShortTitle{Composite Higgs phenomenology}

\author{\speaker{Luigi Del Debbio}\\
  Higgs Centre for Theoretical Physics\\
  The University of Edinburgh \\
  Edinburgh EH9 3FD, Scotland, UK \\
  E-mail: \email{luigi.del.debbio@ed.ac.uk}}

\author{Christoph Englert\\
        School of Physics and Astronomy\\
        The University of
   Glasgow\\Glasgow, G12 8QQ, Scotland, UK\\
        E-mail: \email{christoph.englert@glasgow.ac.uk}}

\author{Roman Zwicky\\
  Higgs Centre for Theoretical Physics\\
  The University of Edinburgh \\
  Edinburgh EH9 3FD, Scotland, UK \\
  E-mail: \email{roman.zwicky@ed.ac.uk}}

\abstract{Several UV complete models of physics beyond the Standard
  Model are currently under scrutiny, their low-energy dynamics being
  compared with the experimental data from the LHC. Lattice
  simulations can play a role in these studies by providing a first
  principles computations of the low-energy constants that describe
  this low-energy dynamics. In this work, we study in detail a
  specific model recently proposed by
  Ferretti~\cite{Ferretti:2014qta}, and discuss the potential impact
  of lattice calculations. }

\FullConference{34th annual International Symposium on Lattice Field Theory\\
		24-30 July 2016\\
		University of Southampton, UK}

\begin{document}

\section{Introduction}
\label{sec:intro}

Strongly-interacting dynamics is one of the possible scenarios
describing the breaking of the electro-weak (EW) symmetry observed in
the Standard Model (SM): new physics beyond the Standard Model (BSM)
is the manifestation of a new theory in a nonperturbative regime,
giving rise to models known as composite Higgs models. The dynamics of
the low-lying states in the spectrum of the strongly-interacting
theory is described by effective field theories (EFT), where the
nonperturbative dynamics is encoded in a number of low-energy
constants (LECs). EFTs are the preferred tool to compare
strongly-interacting models with experimental data. Measurements
translate directly into constraints on the LECs, without ever having
to specify the details of the underlying theory, which is often
referred to as the UV completion. Traditionally, UV completions are
supposed to be gauge theories coupled to matter in one, or several,
representations of the gauge group. We shall refer to the charge under
this new gauge group as {\em hypercolor}, while we will use the word
{\em color} for the usual QCD sector of the SM.

On the other hand, given a UV completion in the form of a hypercolor
theory, first-principle information on the spectrum, and the LECs of
the EFT, can be determined from lattice Monte Carlo simulations of the
UV completion. These simulations are computationally expensive, and it
is desirable to determine the relevant observables, and their desired
precision, from phenomenological studies.

In this work we address the question of the potential impact of
lattice simulations within the particular model presented by Ferretti
in Ref.~\cite{Ferretti:2014qta}. We write down the EFT that encodes
the pattern of spontaneous symmetry breaking in the UV completion,
and analyse the bounds on the LECs from current LHC data. While
similar studies have already been performed, our analysis focusses
especially on finding the experimental bounds on the quantities that
can be computed in lattice simulations.

\section{Low-energy spectrum}
\label{sec:lowenergy}

Let us briefly recall the low-energy description of the model in
Ref.~\cite{Ferretti:2014qta}. The effective action includes the nonlinear
sigma model action describing the Nambu-Goldstone bosons (NGB) of the
strong theory, and their self-interactions, and terms that describe their
coupling to the SM particles. The UV completion is an SU(4) gauge theory,
with 5 Weyl fermions ($\psi^I_{mn}$) in the two-index antisymmetric
representation of the hypercolor group, and 3 Dirac fermions (represented
as a pair of Weyl fermions $\chi^{a}_m$, $\bar{\chi}^{a'}_m$) in the
fundamental representation of the hypercolor group. In the expressions
above, the flavor indices are $I=1,\dots,5$, $a=1,\ldots,3$, and
$a'=1,\ldots,3$; the indices $m,n=1,\dots, 4$ are hypercolor indices. The
theory is symmetric under global transformations belonging to
\begin{equation}
  \label{eq:SymUnbrok}
  G_F = \su{5}\times\su{3}\times\su{3}'\times\gu{1}_X\times\gu{1}'\, .
\end{equation}
The charges of the matter fields under the various symmetry
transformations are explained in detail in
Ref.~\cite{Ferretti:2014qta}. The expected spontaneous symmetry
breaking pattern of the model takes the form
\begin{equation}
  \label{eq:SymBreakPat}
  G_F/H_F = \left(\frac{\su{5}}{\so{5}}\right) \times
  \left(
    \frac{\su{3} \times \su{3}^\prime}{\su{3}_c}
  \right) \times \left(\frac{\gu{1}_X\times \gu{1}^\prime}{\gu{1}_X}\right)\, , 
\end{equation}
induced by the condensates $\langle \epsilon^{mnpq} \psi^I_{mn}
\psi^J_{pq}\rangle \propto \delta^{IJ}$, and $\langle \bar{\chi}^{a'}_m
\chi^a_m\rangle \propto \delta^{a'a}$.

The unbroken subgroup $H_F=\so{5}\times \su{3}_c \times \gu{1}_X$ must
contain the SM group. The EW gauge group $\su{2}_L \times \gu{1}_Y$ is
embedded in the unbroken $\so{5}$ by considering the subgroup
$\so{4} \simeq \su{2}_L \times \su{2}_R$, and then identifying a
$\gu{1}$ subgroup generated by $T^3_R$, the third generator of
$\su{2}_R$, and setting $Y=T^3_R + X$. The unbroken vector subgroup
$\su{3}_c$ is gauged, and identified with the QCD subgroup of the SM.

The 14 NGB in the coset $\su{5}/\so{5}$ can be classified according to
their SM $\su{2}_L\times \gu{1}_R$ charges:
\begin{equation}
  \label{eq:SO5NGB}
  \mathbf{14} \rightarrow \mathbf{1}_0 + \mathbf{2}_{\pm 1/2} + 
  \mathbf{3}_0 + \mathbf{3}_{\pm 1} = \left(\eta, H, \Phi_0, \Phi_\pm\right)\, .
\end{equation}
In composite models the Higgs boson is the NGB denoted by $H$ in the
list above. The field $H$ is a doublet under $\su{2}_L$, and can be
written as a two-component complex field $H=(H_+,H_0)$. If we consider
the hypercolor theory with massless fermions in isolation, the NGB are
exactly massless. The potential for the $H$ field, and more generally
the mass of all NGBs, is generated by the coupling of the hypercolor
theory to the SM fields.  Note that the NGB in the
$\su{3}\times\su{3}'/\su{3}_c$ coset are color charged massless states
that characterize this particular hypercolor theory. We will briefly discuss
their properties below.

The spin 1/2, color triplet states of the hypercolor theory are
natural candidates to play the role of top partners, and hence
generate the mass of the heaviest quark. It is often assumed, when
building EFTs, that there exists one such state of mass $M$, which is
lighter than the typical scale of the hypercolor theory,
$\Lambda_\mathrm{HC}$. A lattice study of the spectrum of the
hypercolor theory would yield unambiguous predictions about the
spectrum of expected resonances. In the context of an EFT description
of the low-energy dynamics, the top partner is introduced in the
effective action as a Dirac fermion field $\Psi$ transforming in the
$(\mathbf{5},\mathbf{3})_{2/3}$ representation of $H_F$. The SM
quantum numbers of this field can be found by decomposing the
$(\mathbf{5},\mathbf{3})_{2/3}$ representation into irreducible
representations of $G_\mathrm{SM} \subset H_F$:
\begin{equation}
  \label{eq:TopPartDecomp}
  (\mathbf{5},\mathbf{3})_{2/3} \rightarrow 
  (\mathbf{3},\mathbf{2})_{7/6} +
  (\mathbf{3},\mathbf{2})_{1/6} +
  (\mathbf{3},\mathbf{1})_{2/3}\, ,
\end{equation}
where the numbers on the RHS denote the irreducible representations of
$\su{3}_c\times\su{2}_L\times\gu{1}_Y$.

As usual the NGB are combined in a field
$\Sigma=\exp\left(\frac{i\Pi}{f}\right)$, where $f$ is the NGB decay
constant of the hypercolor theory, and hence one of the LEC in the
effective action. $\Sigma$ transforms non-linearly under global
$\su{5}$ transformations. It is convenient to write the effective
action in terms of the field $U=\Sigma \Sigma^T$, which transforms
according to $U \mapsto g U g^T$ for $g \in \su{5}$. The coupling to
the SM gauge bosons is obtained by promoting the ordinary derivatives
in the usual chiral lagrangian to covariant derivatives:
\begin{equation}
  \label{eq:EffLagDU}
  \mathcal{L} \supset \frac{f^2}{16} \, \mathrm{tr} \left[
    \left(D_\mu U\right)^\dagger D^\mu U
  \right]\, ,
\end{equation}
where 
\begin{equation}
  \label{eq:CovDer}
  D_\mu U = \partial_\mu U - ig W^a_\mu \left[T^a_L, U\right] - ig' B_\mu
  \left[T^3_R, U\right]\, .
\end{equation}
The mass term for the fermions and the coupling to the SM fermions is
\begin{equation}
  \label{eq:EffLagFerm}
  \mathcal{L} \supset M \bar{\Psi}\Psi + \lambda_q f \bar{\hat
    q}_L \Sigma \Psi_R + \lambda_t f \bar{\hat t}_R \Sigma^* \Psi_L \, ,
\end{equation}
where $\hat{q}_L$ and $\hat{t}_R$ are spurionic embedding of the SM
quarks in the $\mathbf{5}$ and $\mathbf{\bar{5}}$ representations of
$\su{5}$ respectively. The mass of the hypercolor state $M$ is another
LEC that can be determined from lattice studies of the spectrum of the
theory. In this preliminary study, we will simply scan over a sensible
range for $M$. Likewise $\lambda_q$ and $\lambda_t$ are LECs that
determine the mass of the top quark.

The contributions of the SM particles to the Coleman-Weinberg
potential of the NGBs are responsible for the misalignment of the
vacuum, which leads to EW symmetry breaking. In particular only the
fermionic couplings are responsible for negative contributions to the
potential, which are necessary to generate a non-vanishing {\em vev}\
for the $H_0$ component. Following the notation in
Ref.~\cite{Ferretti:2014qta}, we set $H_0=h/\sqrt{2}$, and all other
fields to zero. The couplings of the field $h$ to the SM gauge bosons,
and fermions are as follows:
\begin{eqnarray}
  \mathrm{tr} \left[ U(h) W_\mu U(h)^\dagger W_\mu\right] 
  &=& \frac12
      \left[ 1 +
      \cos(2h/f) \right] W^c_\mu W^c_\mu \, , \\
  \bar{\hat{q}}_L U(h) \hat{t}_R + \bar{\hat{t}}_R U(h)^* \hat{q}_L
  &=&
      \frac{1}{\sqrt{2}} \sin(2h/f)
      \left(
      \bar{t}_L t_R +\bar{t}_R t_L 
      \right)\, .
\end{eqnarray}
The Coleman-Weinberg potential is parametrized by two LECs, $\alpha$
and $\beta$:
\begin{equation}
  \label{eq:CWpotent}
  V(h) \propto \alpha \cos(2h/f) - \beta \sin^2(2h/f)\, .
\end{equation}
The LECs encode the contributions of the SM sector to the
potential, in a way that is analogous to the electromagnetic corrections to
the pion mass~\cite{Das:1967it}. They can be computed from field correlators as
described in Ref.~\cite{Golterman:2015zwa}:
\begin{eqnarray}
  \label{eq:alphabeta}
   \alpha &=& -C_{LR} \frac{1}{2} (3g^2+g'^2) < 0 \, , \\
  2 \beta &=& -  y^2 C_\mathrm{top} \, .
\end{eqnarray}
Clearly EW symmetry breaking can only occur if $\alpha +
2\beta>0$. Moreover the value of the Higgs {\em vev} in units of $f$
is determined by these constants. The LECs in the equations above can
be computed from first principles from correlators in the UV complete
theory. For instance
\begin{equation}
  \label{eq:CLR}
  C_{LR} = \frac{3}{(4\pi)^2} \int_0^\infty dq^2\, q^2 \,
  \Pi_{LR}(q^2) + \mathcal{O}(g^4,g'^4)\, ,
\end{equation}
where $\Pi_{LR}$ is defined from the Lorentz decomposition of  the
current/current correlators: 
\begin{equation}
  \label{eq:PiLR}
  \left( q^2 g_{\mu\nu} - q_\mu q_\nu \right) \Pi_{LR}(q^2)  = 
  \int d^D\!x\, e^{ikx} \mathrm{tr}\langle J_\mu^R(x) J_\nu^L(0) \rangle\, , 
\end{equation}
and
$J^{R,L}_\mu(x)=\bar{\psi}\gamma_\mu \frac{1\pm\gamma_5}{2}
\psi(x)$. Similarly, $C_\mathrm{top}$ can be computed on the lattice,
and therefore the issue of EW dynamical symmetry breaking can be
resolved by numerical simulations.

Finally note that diagonalising the mass term in the effective action
yields the top mass, at leading order in $v$,
\begin{equation}
  \label{eq:mtop}
  m_t = \frac{\sqrt{2} M f \lambda_q \lambda_t}{\sqrt{M^2 + \lambda_q^2 f^2}
    \sqrt{M^2 + \lambda_t^2 f^2}} v \, ,
\end{equation}
or, equivalently,
\begin{equation}
  \label{eq:mtopscaled}
  m_t/v = \frac{\sqrt{2} \rho_M \lambda_q \lambda_t}{\sqrt{1 + \lambda_q^2
      \rho_M^2}\sqrt{1 + \lambda_t^2 \rho_M^2}}\, ,
\end{equation}
where we have introduced the ratio $\rho_M=f/M$, which can be extracted from
numerical simulations. 

Introducing spurions $\hat{q}^i_L \in \mathbf{24}$,
$\hat{u}^i_R \in \mathbf{10}$, and
$\hat{d}^i_R \in \mathbf{\overline{10}}$, where $i=1,2,3$ denotes the quark
family, a mass term for the bottom quark can be introduced
\begin{equation}
  \label{eq:mbterm}
  \mathcal{L} \supset \sqrt{2} \mu_b \mathrm{tr}\left( \bar{\hat{q}}^3_L U
    \hat{d}^3_R + \mathrm{h.c.} \right)\, ;
\end{equation}
the new LEC, $\mu_b$, is determined by requiring that the correct
value is recovered for the bottom quark mass.

\section{Constraints from data}
\label{sec:datacon}

Expanding the exponential in $U$, the effective lagrangian for the
Higgs field reduces to the so-called minimal composite Higgs model
(MCHM)~\cite{Contino:2003ve,Agashe:2004rs,Contino:2006qr}. The Higgs
coupling to the EW gauge bosons is rescaled by a factor
$\sqrt{1-\xi}$ compared to the SM values, where $\xi=v^2/f^2$. Bounds
on $\xi$ from current experimental data have been studied in detail
in the context of MCHM, see e.g.~\cite{Aad:2015pla}.

As discussed in the previous section, this specific model predicts
additional PNGBs, i.e. additional charged Higgs particles, whose
masses are set by the EW contributions to the Coleman-Weinberg
potential.  The color octet of hyper-pions complete the low-energy
spectrum of the theory, the lower bound on their mass being currently
in the multi-TeV regime~\cite{Khachatryan:2014lpa,ATLAS:2016sfd}.

Assuming {\it e.g.}\ a mass $m\approx 200~\mathrm{GeV}$ for the exotic
Higgses, and taking into account the experimental constraints on
$\xi$, the production cross sections for these particles can be
computed. Even for the lightest values of $m$, we find production
cross sections of the order of 100~fb, which decrease exponentially
for higher masses. This is reassuringly consistent with the fact that
these states have not been observed experimentally.

\begin{figure}
  \centering
  \includegraphics[width=.6\textwidth]{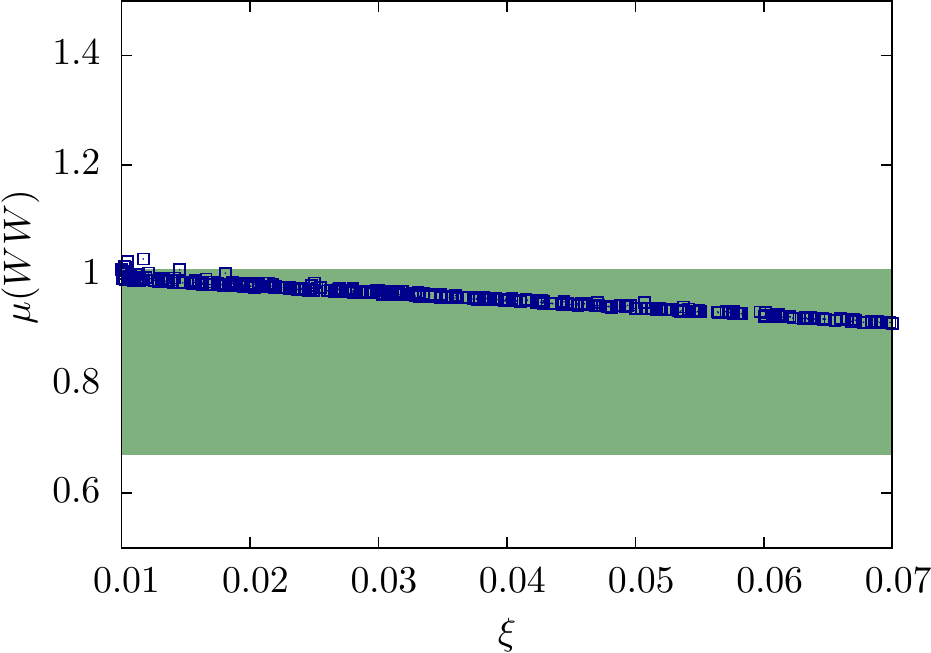}
  \caption{\small Signal strength $\mu(WW)$ for the decay into $h \to WW$ gauge
    bosons. The green band is the error in the LHC data, while the blue data
    results from the scan of the parameters of the EFT. }
  \label{fig:fig1}
\end{figure}

\begin{figure}
  \centering 
  \includegraphics[width=.6\textwidth]{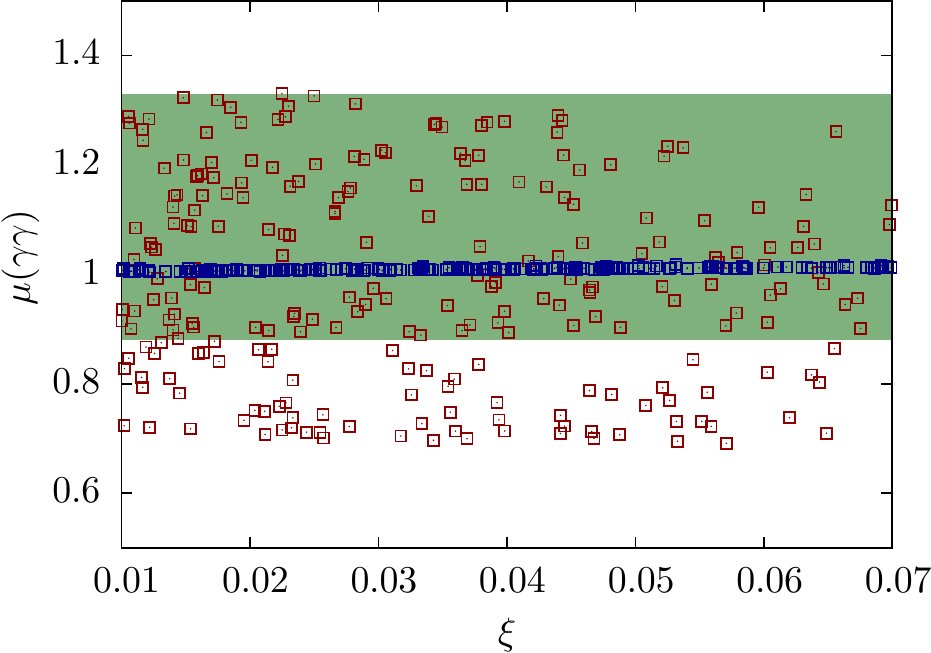}
  \caption{\small Signal strength $\mu(\gamma\gamma)$ for the decay into
    $h \to \gamma\gamma$ gauge bosons. The green band is the error in the
    LHC data, while the blue data
    results from the scan of the parameters of the EFT. The red data includes charged Higgs contributions
    to the diphoton decay width.}
  \label{fig:fig2}
\end{figure}

\begin{figure}
  \centering 
  \includegraphics[width=.6\textwidth]{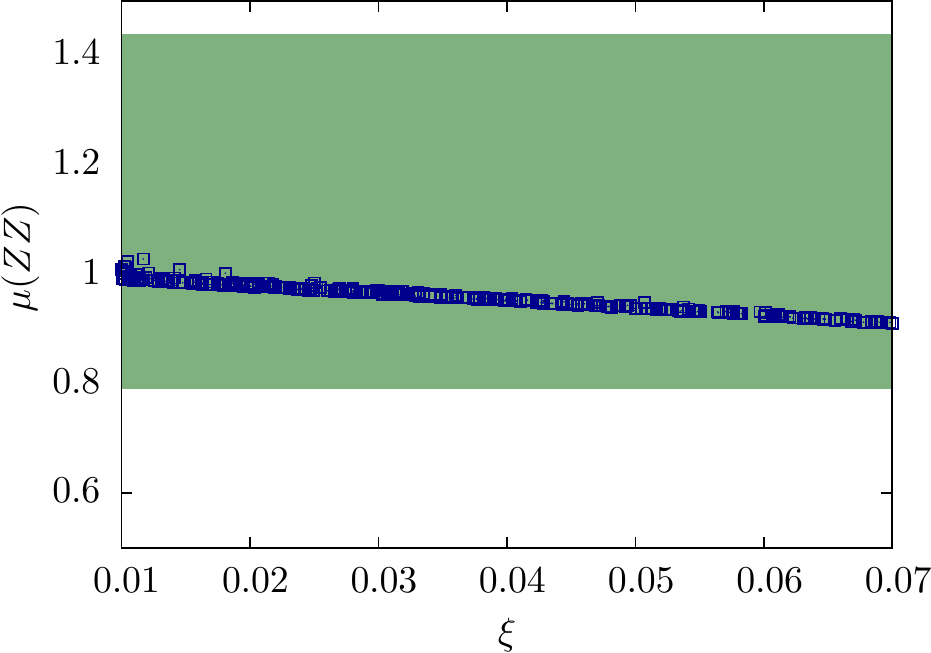}
  \caption{\small Signal strength $\mu(ZZ)$ for the decay into $h \to ZZ$ gauge
    bosons. The green band is the error in the LHC data, while the blue data
    results from the scan of the parameters of the EFT.}
  \label{fig:fig3}
\end{figure}

The customary observable for studies of BSM phenomenology is the
signal strength in a given channel, which is defined as the ratio of
the expected events in the BSM extended model to the events predicted
by the SM.  The additional Higgs particles discussed above modify the
$h \to WW, \gamma\gamma, ZZ$ signal strengths. For this preliminary
report we focus on these three channels only, as examples of possible
applications. In each channel we compute the signal strength predicted
by the effective theory above. The results of our study are shown in
Figs.~\ref{fig:fig1},~\ref{fig:fig2} and \ref{fig:fig3}. In all plots
the LHC data are displayed by the green band as a function of $\xi$,
or, equivalently, $f$. The blue band is obtained by scanning over the
values of the LECs. More precisely, we vary $M$ in the range
$[1.5,3.5]~\mathrm{TeV}$, $\lambda_t$ in the range $[0,4\pi]$. The
value of $\lambda_q$ is then fixed by the mass of the top quark, as
shown in Eq.~\ref{eq:mtopscaled}. Finally $\mu_b$ is engineered to
reproduce the correct bottom quark mass of 4.7 GeV.

The plots show a bit of tension between the signal strength
meaurements by ATLAS and CMS and the prediction in the model. 
In
particular, $WW$ production is currently favoured slightly below the SM expectation.
However, the presence of the newly charged scalar states introduces additional
freedom to obtain a signal strength in the diphoton channel that is
comparable to the observed limits over a wide range of signal strengths. The width of the blue and red bands
highlights the sensitivity of the Higgs branching to photon pairs. 
%On
%the other hand, since $h\to \gamma \gamma$ is only a small partial
%decay width the impact on the prompt $h\to ZZ, WW$ decays is only
%small. Hence, the size of the blue band (signalling the scan's
%spread) in these channels is a lot more narrow.
 
\section{Conclusions}
\label{sec:concl}

The preliminary results of our study suggest that the UV theory
considered here is compatible with the data analysed so far by the
experiments at the LHC, for a sensible range of the couplings in the
effective action, despite showing some tension e.g. the $h \to WW$. Currently, the main
source of uncertainty is the experimental error in the data, while the
variations in the theoretical predictions due to the scan in the space
of parameters seems to be rather smaller.

We have identified a number of LECs that appear in the effective
action, and have discussed the possibility of measuring them in
lattice simulations of the UV complete underlying theory. Some of the
LEC like the condensates, $C_{LR}$, $C_\mathrm{top}$ allow us to check
the pattern of symmetry breaking, and the generation of the correct
Higgs potential. Note that a robust prediction of the sign of the
combination $\alpha+2\beta$ would already yield valuable
information. Lattice results would also constrain the region that
needs to be explored when scanning over the space of LECs. In
particular a determination of $\xi$ and $\rho_M$ would reduce the scan
to a lower-dimensional subspace, and therefore could constrain the
model substantially. 

As the experimental errors decrease, lattice simulations can lead to
stringent tests of BSM models with a UV completion.

\end{document}